\documentclass[letter,oneside,9pt,twocolumn,article]{aiaa}

\usepackage{rotating}
\usepackage{bm}
\usepackage{flushend}
\usepackage[top=0.5in, bottom=1.0in, left=0.7in, right=0.7in]{geometry}
\usepackage[bf,footnotesize]{subfigure}
\usepackage{adjustbox}
\usepackage{tabularx}
\usepackage{multirow}
\usepackage{multicol}
\usepackage{indentfirst}
\usepackage[hidelinks]{hyperref}
\usepackage{float}
\usepackage{xcolor}
\usepackage[super,comma,sort,compress]{natbib}
\usepackage{mwe}
\usepackage[markcase=noupper]{scrlayer-scrpage}
\usepackage{caption}

\ihead*{Submitted for publication in the Journal of Chemical Physics}
\ohead*{\pagemark}
\ifoot*{}
\ofoot*{\rm Unrestricted content}
\renewcommand{\fontsizetable}{\footnotesize\scalefont{1.0}}

\author{
Colby Cinqmars\thanks{Graduate student, colbycinqmars@arizona.edu.} ~and
Bernard Parent\thanks{Associate Professor, bparent@arizona.edu. Author to whom correspondence should be addressed.}
\\[0.3em] \it University of Arizona, Tucson, AZ 85721, USA.
}

\title{Analytic Model for Transient Plateau Filling in Vibrational Distributions of Periodically Driven Plasmas}

\abstract{
In periodically driven molecular plasmas, the vibrational distribution function is governed by the competition between vibration--vibration (V--V) up-pumping along the anharmonic ladder and vibration--translation (V--T) relaxation, producing the Treanor--Gordiets distribution. When vibrational energy accumulates across cycles faster than V--V exchange can populate the upper levels, the Gordiets plateau remains transiently depleted. Existing algebraic models assume a quasi-steady plateau, whereas exact transient solutions of the nonlinear V--V diffusion equation require the elapsed time since excitation and the source strength, which are unavailable in multidimensional simulations that track only local macroscopic variables. Here, elapsed time is replaced by a local vibrational energy timescale formed from the vibrational energy per molecule and its period-averaged rate of change. This reduces the diffusion equation in level space to an ordinary differential equation solved in closed form without linearization, yielding a level-dependent correction to the steady plateau governed by a single dimensionless group: the number of V--V collisions per energy timescale weighted by the plateau amplitude at the Treanor minimum. The distribution front advances as the square root of this group, and the steady distribution is recovered as the group grows large. Compared with state-to-state master-equation calculations for N$_2$ under nanosecond pulse trains spanning 300--1500~K, 0.041--1~atm, and 1--100~kHz (1440 sampled states), the model reduces the 90th-percentile error in level populations from 1.88~decades for the steady model and 1.32~decades for an adapted self-similar solution to 0.53~decades, at essentially the evaluation cost of the steady model.
}

\nomenclature{
\begin{nomenclaturelist}{Roman symbols}
\item[$A$]{FHO repulsion potential coefficient, eV}
\item[$a$]{V--V diffusion prefactor, dimensionless}
\item[$A_{\rm MW}$]{Millikan--White correlation constant, K$^{1/3}$}
\item[$B_{\rm VT}$]{V--T plateau drain coefficient, dimensionless}
\item[$c$]{speed of light, cm/s}
\item[$C_1$]{ZNS plateau height set by source power, dimensionless}
\item[$D_0$]{dissociation energy, eV}
\item[$e_{\rm v}$]{vibrational energy per molecule above the ground level, J}
\item[$e_{{\rm v},k}$]{$e_{\rm v}$ at the end of the interpulse of cycle $k$, J}
\item[${\mathcal E}_n$]{energy of vibrational level $n$, J}
\item[$f$]{pulse repetition frequency, kHz}
\item[$h$]{Planck constant, J\,s}
\item[$\hbar$]{reduced Planck constant $h/2\pi$, J\,s}
\item[$k$]{pulse cycle index}
\item[$k_{\rm B}$]{Boltzmann constant, J/K}
\item[$k_{\rm VT}^{n_{\rm i}\rightarrow n_{\rm f}}$]{V--T rate coefficient for a molecule going from level $n_{\rm i}$ to level $n_{\rm f}$ (e.g., $k_{\rm VT}^{1\rightarrow0}$, the fundamental rate), cm$^3$/s}
\item[$k_{\rm VV}^{n_{\rm i}\rightarrow n_{\rm f},\,w_{\rm i}\rightarrow w_{\rm f}}$]{V--V rate coefficient for a collision in which one molecule goes from level $n_{\rm i}$ to $n_{\rm f}$ and its partner from $w_{\rm i}$ to $w_{\rm f}$ (e.g., $k_{\rm VV}^{1\rightarrow0,\,0\rightarrow1}$, the fundamental resonant exchange), cm$^3$/s}
\item[$m$]{number of quanta exchanged in a transition, $|n_{\rm f}-n_{\rm i}|$}
\item[$n$]{vibrational quantum number}
\item[$n'$]{dummy integration variable for $n$}
\item[$n^\star$]{Treanor minimum quantum number}
\item[$n_{\rm diss}$]{first unbound vibrational level (dissociation limit)}
\item[$n_{\rm f}$]{final vibrational level of a molecule in a collision}
\item[$n_{\rm front}$]{distribution front level}
\item[$n_{\rm i}$]{initial vibrational level of a molecule in a collision}
\item[$N$]{total number density of the molecule, cm$^{-3}$}
\item[$N_0$]{ground state number density, cm$^{-3}$}
\item[$N_n$]{number density at vibrational level $n$, cm$^{-3}$}
\item[$N_n^{\rm StS}$]{state-to-state reference number density at level $n$, cm$^{-3}$}
\item[$P$]{pressure, Pa}
\item[$q_n$]{scaled population variable $(n+1)N_n/N$, dimensionless}
\item[$R_0$]{FHO maximum interaction distance, m}
\item[$S$]{number of vibrational states tracked in the mixture}
\item[$t$]{time, s}
\item[$T_{\rm e}$]{electron temperature, eV}
\item[$T_{\rm g}$]{translational gas temperature, K}
\item[$T_{\rm LT}$]{Landau--Teller temperature, K}
\item[$T_{{\rm LT},n}$]{level-dependent Landau--Teller temperature, K}
\item[$T_{\rm v}$]{vibrational temperature, K}
\item[$\bar{u}$]{mean relative collision speed, m/s}
\item[$U(n)$]{unsteady plateau correction factor, dimensionless}
\item[$U_{\rm ZNS}(n)$]{ZNS-adapted correction factor, dimensionless}
\item[$w$]{vibrational level of V--V partner molecule}
\item[$w_{\rm f}$]{final vibrational level of V--V partner molecule}
\item[$w_{\rm i}$]{initial vibrational level of V--V partner molecule}
\item[$W(n)$]{dissociation wall factor, dimensionless}
\item[$x_{\rm e}$]{first-order Dunham anharmonicity constant, dimensionless}
\item[$y_{\rm e}$]{second-order Dunham anharmonicity constant, dimensionless}
\end{nomenclaturelist}
\begin{nomenclaturelist}{Greek symbols}
\item[$\alpha$]{FHO intermolecular potential steepness, \AA$^{-1}$}
\item[$\Gamma$]{plateau amplitude at $n^\star$, dimensionless}
\item[$\delta_{\rm VT}$]{V--T rate-defect parameter, dimensionless}
\item[$\delta_{\rm VV}$]{V--V rate-defect parameter, dimensionless}
\item[$\varepsilon_{\rm v}$]{vibrational energy deposited per pulse per molecule, eV}
\item[$\theta_{\rm v}$]{characteristic vibrational temperature of the fundamental transition, $({\mathcal E}_1-{\mathcal E}_0)/k_{\rm B}$, K}
\item[$\Lambda$]{V--V collisions per energy timescale, dimensionless}
\item[$\mu$]{reduced mass of collision pair, kg}
\item[$\mu_{\rm amu}$]{reduced mass of collision pair, amu}
\item[$\nu$]{V--V diffusion coefficient in level space, s$^{-1}$}
\item[$\tau_{\rm e}$]{vibrational energy timescale, s}
\item[$\tau_{\rm VT}$]{V--T relaxation time, s}
\item[$\phi$]{fraction of the plateau filled, dimensionless}
\item[$\omega_{\rm e}$]{harmonic vibrational frequency, cm$^{-1}$}
\item[$\omega_1$]{fundamental vibrational transition frequency $k_{\rm B}\theta_{\rm v}/\hbar$, rad/s}
\item[$\omega_n$]{anharmonic transition frequency for $n\to n-1$, cm$^{-1}$}
\end{nomenclaturelist}
}

\begin{document}
\maketitle

\section{Introduction}
\dropword
Non-equilibrium vibrational distributions in molecular gases arise in
periodically driven plasmas where vibrational energy accumulates from
one cycle to the next, including plasma-assisted combustion,\cite{bak2011,adamovich2015,chen2023,castela2016,dijoud2025}
laser-induced plasmas for flame anchoring,\cite{dumitrache2017,yu2017}
sinusoidal discharges in semiconductor
processing,\cite{tinck2016,foucher2015,annusova2018}
and atmospheric-pressure plasma sources.\cite{jiang2022,miyake2024,bruggeman2016,davies2023}
In all of these settings, predicting energy coupling and chemical
reaction rates requires the molecular vibrational distribution
function (VDF). Although the excitation mechanism varies across
applications, whether via electron impact or laser absorption, the
subsequent redistribution of vibrational quanta is governed by the
same two competing heavy-particle processes: vibration--vibration
(V--V) exchange pumps quanta upward along the anharmonic vibrational
ladder, while vibration--translation (V--T) relaxation drains them
into translational energy. The competition between these processes
produces the characteristic two-branch Treanor--Gordiets
distribution,\cite{treanor1968,gordiets1974} in which V--V exchange
sustains a Treanor distribution at low levels and V--T relaxation
terminates a plateau extending across the upper levels.

Because the Treanor--Gordiets distribution governs both heavy-particle
dissociation rates and electron kinetics---such as superelastic
electron--vibrational heating, for which a unified analytic closure
was recently derived\cite{parent2026}---a substantial body of work has
sought closed-form VDF descriptions that can be evaluated directly
within multidimensional flow and plasma simulations. The most complete
treatment, the state-to-state (StS) method, integrates a coupled rate
equation for every vibrational level to capture the full VDF across all
regimes. However, its computational cost scales as $O(S^3)$
per grid point, where $S$ is the number of tracked vibrational
states, making it prohibitive in multidimensional simulations with
$10^5$ or more grid points. The analytic model of Adamovich, Macheret,
and Rich\cite{adamovich1994} (AMR) overcomes this barrier by extending
the foundational theory of Gordiets et al.\cite{gordiets1974} into a
fully algebraic framework. Rather than integrating differential rate
equations, AMR reconstructs both the Treanor branch and the Gordiets
plateau directly from three local macroscopic quantities: gas
temperature, vibrational temperature, and molecular number density.
This combination of physical fidelity and negligible evaluation cost
makes the AMR model well suited for coupled simulations, and it serves
here as the steady-state foundation of the present work.

The AMR model, however, assumes that the VDF has reached a
quasi-steady state. In the transient filling regime of a periodically
driven plasma, this assumption breaks down: vibrational energy
accumulates across cycles faster than V--V exchange can relay quanta
to the upper levels, leaving the Gordiets plateau depleted relative to
its steady-state profile when the next pulse arrives. Prior work has
approached this transient regime through either numerical continuum formulations or analytical solutions. Continuum
Fokker--Planck formulations, such as the modified approach of Altin et
al.,\cite{altin2023} replace the discrete master equation with a partial differential equation (PDE) in level space, but still require
numerical integration and do not meaningfully reduce the computational
cost relative to StS. Conversely, the analytical theory of Zhdanok, Napartovich, and Starostin\cite{zhdanok1979}
(ZNS) and the related analyses of Gordiets and Zhdanok\cite{gordiets1986} yield
closed-form transient solutions of the nonlinear V--V diffusion equation, with ZNS
giving self-similar solutions for a source switched on or off and for a single
excitation pulse. Yet each
of these analytical solutions requires the elapsed time since the
excitation event as an explicit input, together with either the source
power or the deposited pulse energy. Neither quantity is available in
a multidimensional flow or plasma simulation, which transports only the
local macroscopic state at each grid point (such as gas temperature,
density, and vibrational energy) and retains no record of the
excitation history. Thus, to our knowledge, no analytic model currently
exists for the transient filling regime that can be evaluated solely
from local macroscopic variables.

This work derives such a model and validates it against StS
calculations of $\mathrm{N_2}$ in repetitively pulsed nanosecond
discharges. Building on the steady AMR foundation, the present model
reconstructs the full transient VDF algebraically from four local
macroscopic inputs---gas temperature, molecular number density,
vibrational energy per molecule $e_{\rm v}$, and its period-averaged
rate of change ${\rm d}e_{\rm v}/{\rm d}t$---without ordinary
differential equation (ODE) integration. Specifically, elapsed time in
the ZNS nonlinear diffusion equation is replaced by a local
vibrational energy timescale $\tau_{\rm e} \equiv e_{\rm v}/({\rm d}e_{\rm v}/{\rm d}t)$.
This substitution reduces the governing PDE in level space to an ODE that is solved in closed form while retaining
the full nonlinearity of V--V transport. The resulting solution shows
that transient plateau filling is governed by a single dimensionless
group and that the distribution front advances as its square root,
recovering the steady AMR plateau as this group grows large. Because ${\rm d}e_{\rm v}/{\rm d}t$ is a period-averaged rate available
from the vibrational energy equation of a flow or plasma solver (in a gas at
rest, simply by storing $e_{\rm v}$ from the previous cycle) and is defined
identically regardless of the excitation mechanism, the model is expected to apply broadly to periodically
driven plasmas in which vibrational energy builds across cycles, although this
remains to be tested beyond the N$_2$ cases considered here.

\section{Model Derivation and Implementation}
\label{closureDerivation}
The present model is developed in three steps: the steady-state vibrational
distribution is established first (Sec.~\ref{SteadyState}), then extended to
the transient filling regime (Sec.~\ref{TransientCorrection}), and finally
assembled into an algorithm that computes the distribution from local
macroscopic variables (Sec.~\ref{ModelImplementation}).

\subsection{Steady-State Vibrational Distribution}
\label{SteadyState}
We begin with the steady-state vibrational distribution model of
AMR,\cite{adamovich1994} which divides the vibrational manifold into two
physically distinct regions separated by the Treanor minimum $n^\star$. Below
$n^\star$, near-resonant V--V exchange dominates and drives the populations
toward the Treanor distribution;\cite{treanor1968} above $n^\star$, the
competition between V--V up-pumping and V--T relaxation produces the Gordiets
plateau.\cite{gordiets1974,gordiets1986,gordiets1988}

There, the populations follow the Treanor distribution\cite{treanor1968,parent2026}
\begin{equation}
    \frac{N_n}{N_0} = \exp\!\left[
        -\frac{n\theta_{\rm v}}{T_{\rm v}}
        + \frac{n\theta_{\rm v}-({\mathcal E}_n-{\mathcal E}_0)/k_{\rm B}}{T_{\rm g}}
    \right] \qquad \text{for } n \leq n^\star,
    \label{eq:treanor}
\end{equation}
where $N_0$ is the ground-state number density, $N_{n}$ is the number density of
vibrational level $n$, $k_{\rm B}$ is the Boltzmann constant, $T_{\rm g}$ is the
translational temperature, and $T_{\rm v}$ is the vibrational temperature
defined by Treanor\cite{treanor1968} as
$N_1/N_0 = \exp(-\theta_{\rm v}/T_{\rm v})$, with
$\theta_{\rm v} \equiv ({\mathcal E}_1-{\mathcal E}_0)/k_{\rm B}$ denoting the
characteristic vibrational temperature of the fundamental
transition.\cite{parent2026} The vibrational energy levels ${\mathcal E}_n$ are
given by the second-order Dunham expansion\cite{parent2026}
\begin{equation}
    {\mathcal E}_n = hc\,\omega_{\rm e}\!\left[(n+\tfrac{1}{2})
      - x_{\rm e}(n+\tfrac{1}{2})^2
      + y_{\rm e}(n+\tfrac{1}{2})^3\right],
    \label{eq:dunham}
\end{equation}
where $h$ is the Planck constant, $c$ is the speed of light, $\omega_{\rm e}$ is
the harmonic vibrational frequency, and $x_{\rm e}$ and $y_{\rm e}$ are the
dimensionless first- and second-order anharmonicity constants obtained from the
spectroscopic constants $\omega_{\rm e} x_{\rm e}$ and $\omega_{\rm e} y_{\rm e}$
in $\text{cm}^{-1}$.\cite{nistwebbook} Evaluating Eq.~\eqref{eq:dunham} at $n=0$
and $n=1$ relates the characteristic vibrational temperature to the
spectroscopic constants via
$k_{\rm B}\theta_{\rm v} = hc\,\omega_{\rm e}(1-2x_{\rm e}+3.25\,y_{\rm e})$.
Equation~\eqref{eq:dunham} also defines the first unbound vibrational level
$n_{\rm diss}$ (the dissociation wall) as the smallest integer $n$ satisfying
${\mathcal E}_n - {\mathcal E}_0 > D_0$, where $D_0$ is the dissociation energy;
for N$_2$, $D_0 = 9.759$~eV\cite{nistwebbook} gives $n_{\rm diss} = 48$,
corresponding to 48 bound vibrational levels ($n=0$ to $47$).

The Treanor minimum $n^\star$ is found by setting the derivative of the Treanor
exponent to zero using the second-order Dunham expansion and selecting the
physical root as outlined by Parent:\cite{parent2026}
\begin{equation}
    n^\star = \frac{2x_{\rm e}-3y_{\rm e}-\sqrt{\,4x_{\rm e}^2-12y_{\rm e}\!\left[\dfrac{T_{\rm g}}{T_{\rm v}}
          +(2x_{\rm e}-3.25\,y_{\rm e})\!\left(1-\dfrac{T_{\rm g}}{T_{\rm v}}\right)\right]}}{6y_{\rm e}}.
    \label{eq:nstar}
\end{equation}
For $y_{\rm e}<0$, as for N$_2$, $n^\star$ is always real and positive. For
$y_{\rm e}>0$, the square root can become imaginary (e.g., for O$_2$ when
$T_{\rm g}/T_{\rm v}$ exceeds about 0.63), in which case the linear Treanor
minimum, $n^\star = T_{\rm g}/(2x_{\rm e}T_{\rm v}) - T_{\rm g}/T_{\rm v} + 1/2$,
obtained from Eq.~\eqref{eq:nstar} as $y_{\rm e}\to0$,\cite{parent2026} is used
instead. When $T_{\rm g}/T_{\rm v}$ exceeds a threshold, $n^\star$ surpasses
$n_{\rm diss}$ and the entire bound manifold follows the Treanor distribution. In
practice, $n^\star$ is not rounded, and the Treanor branch is evaluated at the
integer levels $n \leq n^\star$.

Above $n^\star$, the steady distribution follows from Adamovich et
al.,\cite{adamovich1994} with the radiative term absent for a homonuclear
molecule and the long-range V--V correction neglected (the latter arises from
multipole interactions and is significant only for polar species):
\begin{equation}
    \frac{N_n}{N} =
    \frac{\Gamma\,W(n)}{n+1}
    - \frac{B_{\rm VT}\,e^{\delta_{\rm VT}n}}{n+1}
    \qquad \text{for } n > n^\star,
    \label{eq:plateau}
\end{equation}
where $N$ is the total number density of the molecule. The first term is the
classical Gordiets plateau sustained by V--V exchange and the second is the
V--T relaxation drain. Because the V--T term grows exponentially with $n$ and
eventually exceeds the plateau term, all levels above this cutoff point are set
to zero.\cite{adamovich1994} The plateau amplitude $\Gamma$ is set by matching
to the Treanor branch at $n^\star$:\cite{adamovich1994}
\begin{equation}
    \Gamma = (n^\star+1)\,\frac{N_{n^\star}}{N},
    \label{eq:Gamma}
\end{equation}
where $N_{n^\star}$ is evaluated from Eq.~\eqref{eq:treanor}. A small
discontinuity between the Treanor and plateau branches may appear at $n^\star$
because the two expressions are matched in amplitude through $\Gamma$ but not in
slope. Because Eq.~\eqref{eq:treanor} gives the populations relative to $N_0$
whereas Eqs.~\eqref{eq:plateau} and~\eqref{eq:Gamma} are normalized by $N$, the
normalization constant $N_0$ couples the two branches and is obtained by summing
the distribution over all bound levels $n$ and enforcing conservation of
molecule number. Following Macheret,\cite{macheret1985} the wall factor $W(n)$
accounts for the loss of molecules pumped past the dissociation boundary by V--V
exchange,\cite{adamovich1994} assuming that molecules reaching the boundary are
removed faster than resonant V--V exchange can supply them (Eq.~21 of
Ref.~\citenum{adamovich1994}):
\begin{equation}
    W(n) = \left(\frac{n_{\rm diss} - n}{n_{\rm diss} - n^\star}\right)^{1/2}.
    \label{eq:wall}
\end{equation}

The V--T coefficient $B_{\rm VT}$ follows from Eq.~16 of Adamovich et
al.\cite{adamovich1994} for a pure diatomic gas without the long-range V--V
correction (i.e., with both the diatomic mole fraction and the long-range
correction factor set to unity):
\begin{equation}
    B_{\rm VT} = \frac{k_{\rm VT}^{1\rightarrow0}}{k_{\rm VV}^{1\rightarrow0,\,0\rightarrow1}}\cdot
    \frac{k_{\rm B}T_{\rm g}\,\delta_{\rm VV}^3}{12\,hc\,\omega_{\rm e}x_{\rm e}\,\delta_{\rm VT}},
    \label{eq:BVT}
\end{equation}
where $\delta_{\rm VV}$ and $\delta_{\rm VT}$ are rate-defect parameters
defined below.

Evaluating Eq.~\eqref{eq:BVT} requires the fundamental rate coefficients
$k_{\rm VV}^{1\rightarrow0,\,0\rightarrow1}$ and $k_{\rm VT}^{1\rightarrow0}$.
A V--V collision involves two molecules exchanging vibrational quanta with rate
coefficient $k_{\rm VV}^{n_{\rm i}\rightarrow n_{\rm f},\,w_{\rm i}\rightarrow w_{\rm f}}$,
where $n_{\rm i}$, $n_{\rm f}$ and $w_{\rm i}$, $w_{\rm f}$ are the initial and
final levels of the colliding pair, whereas a V--T collision transfers
vibrational quanta from a single molecule to translational motion with rate
coefficient $k_{\rm VT}^{n_{\rm i}\rightarrow n_{\rm f}}$. Adamovich et al.\
treated both fundamental rates as empirical inputs inferred from CO experiments
and provided no closed-form expressions. Here,
$k_{\rm VV}^{1\rightarrow0,\,0\rightarrow1}$ is evaluated from the forced
harmonic oscillator--free rotation (FHO--FR) theory of
Adamovich,\cite{adamovich2001} the most complete first-principles V--V rate
model available. The accuracy of the present model is therefore tied to the
domain of validity of FHO--FR, which has been validated across a range of diatomic species.\cite{adamovich2001} The fundamental V--T rate
$k_{\rm VT}^{1\rightarrow0}$ is obtained from the Millikan--White
correlation,\cite{millikan1963systematics} which is widely used and requires
only the reduced mass and characteristic vibrational temperature of the
collision pair, allowing it to be applied to any diatomic species without
species-specific fitting.

For the fundamental resonant exchange, Adamovich gives the thermally averaged
FHO--FR rate coefficient in closed form (Eq.~23 of
Ref.~\citenum{adamovich2001} with $m=1$, where $m=|n_{\rm f}-n_{\rm i}|$ is the
number of quanta exchanged):
\begin{equation}
    k_{\rm VV}^{1\rightarrow0,\,0\rightarrow1} = \frac{3}{64}\,\pi R_0^2\,\bar{u}\,
    \frac{\alpha^2 k_{\rm B} T_{\rm g}}{\mu\,\omega_1^2},
    \label{eq:kvv}
\end{equation}
where $R_0 = \alpha^{-1}\ln(A/k_{\rm B}T_{\rm g})$ is the FHO maximum
interaction distance, with $\alpha$ the intermolecular potential steepness and
$A$ the repulsion coefficient ($4$~\AA$^{-1}$ and $1730$~eV for N$_2$);
$\bar{u} = (8k_{\rm B}T_{\rm g}/\pi\mu)^{1/2}$ is the mean relative collision
speed for reduced mass $\mu$; and $\omega_1 = k_{\rm B}\theta_{\rm v}/\hbar$ is
the fundamental vibrational transition frequency, with $\hbar$ the reduced
Planck constant.

Next, for $k_{\rm VT}^{1\rightarrow0}$, Landau--Teller
theory\cite{nikitin2008,landau1936theorie} relates the relaxation time
$\tau_{\rm VT}$ to the fundamental rate coefficient via
$\tau_{\rm VT}^{-1} = N\cdot k_{\rm VT}^{1\rightarrow0}\cdot
(1 - e^{-\theta_{\rm v}/T_{\rm g}})$. Solving for $k_{\rm VT}^{1\rightarrow0}$
and substituting $N = P/(k_{\rm B}T_{\rm g})$, where $P$ is the pressure, yields
\begin{equation}
    k_{\rm VT}^{1\rightarrow0} =
    \frac{1}{\bigl[(\tau_{\rm VT}\,P)/(k_{\rm B}T_{\rm g})\bigr]
             \bigl(1 - e^{-\theta_{\rm v}/T_{\rm g}}\bigr)},
    \label{eq:kvt}
\end{equation}
in which $P$ cancels between $N$ and $(\tau_{\rm VT}\,P)$, leaving
$k_{\rm VT}^{1\rightarrow0}$ a function of $T_{\rm g}$ alone. The
Millikan--White correlation\cite{millikan1963systematics} gives
$(\tau_{\rm VT}\,P)$ in $\text{atm}\cdot\text{s}$ as
\begin{equation}
    \tau_{\rm VT}\,P =
    \exp\!\left[A_{\rm MW}\!\left(T_{\rm g}^{-1/3}
    - 0.015\,\mu_{\rm amu}^{1/4}\right) - 18.42\right],
    \label{eq:MW}
\end{equation}
which is multiplied by $101\,325\text{ Pa/atm}$ to convert to
$\text{Pa}\cdot\text{s}$ for use in Eq.~\eqref{eq:kvt}. Here,
$A_{\rm MW} = 1.16\times10^{-3}\mu_{\rm amu}^{1/2}(hc\,\omega_{\rm e}/k_{\rm B})^{4/3}$
(Eq.~3a of Ref.~\citenum{millikan1963systematics}), where the characteristic
oscillator temperature is taken, as in Millikan and White, as the harmonic value
$hc\,\omega_{\rm e}/k_{\rm B}$ (3395~K for N$_2$ in their Table~I), and
$\mu_{\rm amu}$ is the reduced mass of the collision pair in atomic mass units
($14$~amu for N$_2$--N$_2$).

It remains to define the rate-defect parameters $\delta_{\rm VV}$ and
$\delta_{\rm VT}$ in Eqs.~\eqref{eq:plateau} and~\eqref{eq:BVT}. The parameter
$\delta_{\rm VV}$ enters the AMR V--V rate model,\cite{adamovich1994} which
gives the single-quantum exchange rate between a molecule at level $n$ and a
partner at level $w$ as
\begin{align}
    k_{\rm VV}^{n+1\rightarrow n,\,w\rightarrow w+1} &= (n+1)(w+1)\,k_{\rm VV}^{1\rightarrow0,\,0\rightarrow1}
    \nonumber\\
    &\quad\times
    e^{-\delta_{\rm VV}|n-w|}\!\left[\tfrac{3}{2}
    - \tfrac{1}{2}e^{-\delta_{\rm VV}|n-w|}\right].
    \label{eq:kvv_nm}
\end{align}
When $n$ and $w$ are far apart, the energy mismatch of the non-resonant exchange
must be absorbed by translation, suppressing the rate exponentially with
increasing $|n-w|$. Following Gordiets and Zhdanok,\cite{gordiets1986} we use
\begin{equation}
    \delta_{\rm VV} = \frac{0.427}{\alpha}\sqrt{\frac{\mu_{\rm amu}}{T_{\rm g}}}\,
    \frac{hc\,\omega_{\rm e}x_{\rm e}}{k_{\rm B}},
    \label{eq:delta_vv}
\end{equation}
with $\alpha$ in \AA$^{-1}$, $\mu_{\rm amu}$ in amu, and
$hc\,\omega_{\rm e}x_{\rm e}/k_{\rm B}$ and $T_{\rm g}$ in K.

Likewise, the parameter $\delta_{\rm VT}$ governs the exponential growth of the
V--T rate with vibrational level,\cite{adamovich1994}
$k_{\rm VT}^{n\rightarrow n-1} = n\cdot k_{\rm VT}^{1\rightarrow0}\cdot
e^{\delta_{\rm VT}(n-1)}$, where the prefactor $n$ is the harmonic level
scaling.\cite{nikitin2008} Because Adamovich et al.\ give no prescribed formula
for $\delta_{\rm VT}$, it is derived here from the Landau--Teller
law,\cite{nikitin2008}
$k_{\rm VT}^{1\rightarrow0} \propto \exp[-3(T_{\rm LT}/T_{\rm g})^{1/3}]$, where
the Landau--Teller temperature scales with the square of the transition
frequency as $T_{\rm LT} = 2\pi^4 c^2\omega_{\rm e}^2\mu/(\alpha^2 k_{\rm B}) \propto \omega_{\rm e}^2$.
Anharmonicity reduces the transition frequency for the $n \to n{-}1$ transition
to $\omega_n = \omega_{\rm e}(1-2x_{\rm e}n)$,\cite{parent2026} yielding a
level-dependent $T_{{\rm LT},n} = T_{\rm LT}(1-2x_{\rm e}n)^2$. Substituting
$T_{{\rm LT},n}$ and expanding the exponent to first order in $x_{\rm e}$ gives
\begin{equation}
    -3\!\left(\frac{T_{{\rm LT},n}}{T_{\rm g}}\right)^{\!1/3}
    = -3\!\left(\frac{T_{\rm LT}}{T_{\rm g}}\right)^{\!1/3}
    + 4x_{\rm e}\!\left(\frac{T_{\rm LT}}{T_{\rm g}}\right)^{\!1/3}\!(n-1).
    \label{eq:LT_expanded}
\end{equation}
The first term on the right-hand side is already contained in
$k_{\rm VT}^{1\rightarrow0}$, while the second multiplies the rate by
$\exp[4x_{\rm e}(T_{\rm LT}/T_{\rm g})^{1/3}(n-1)]$. Combining this factor with
the harmonic scaling\cite{nikitin2008} gives
\begin{equation}
    k_{\rm VT}^{n\rightarrow n-1} = n\cdot k_{\rm VT}^{1\rightarrow0}\cdot
    \exp\!\left[4x_{\rm e}\!\left(\frac{T_{\rm LT}}{T_{\rm g}}
    \right)^{\!1/3}\!(n-1)\right],
    \label{eq:kvt_scaled}
\end{equation}
from which
\begin{equation}
    \delta_{\rm VT} = \frac{4}{3} x_{\rm e} A_{\rm MW} T_{\rm g}^{-1/3},
    \label{eq:delta_vt}
\end{equation}
where $T_{\rm LT}$ has been expressed through the Millikan--White correlation as
$T_{\rm LT} = (A_{\rm MW}/3)^3$. For N$_2$, Eq.~\eqref{eq:delta_vt} gives
$\delta_{\rm VT}\approx0.27$ at 300~K and $0.16$ at 1500~K, which is of the same
order as $\delta_{\rm VV}$ ($\approx0.48$ and $0.21$) and consistent with the
approximation $\delta_{\rm VT}\approx\delta_{\rm VV}$ adopted by ZNS for a pure
diatomic gas;\cite{zhdanok1979} the two parameters nonetheless differ in their
temperature scaling ($T_{\rm g}^{-1/3}$ versus $T_{\rm g}^{-1/2}$), and the
Landau--Teller form is retained here. Equations~\eqref{eq:treanor}--\eqref{eq:delta_vt}
constitute the complete quasi-steady adjusted AMR model, which we now extend to
the transient filling regime.

\subsection{Transient Correction}
\label{TransientCorrection}
In the steady plateau of Eq.~\eqref{eq:plateau}, V--V up-pumping supplies quanta
across all levels $n > n^\star$ at a settled rate while V--T relaxation locally
drains them and dissociation removes molecules at $n_{\rm diss}$. During the
transient filling regime of a repetitively pulsed discharge, however, V--V
exchange must relay quanta upward from $n^\star$ level by level, so the upper
plateau lags behind its quasi-steady form. Because V--T relaxation acts locally
at each level---depending only on the instantaneous population $N_n$ and
$T_{\rm g}$, not on the time history of the distribution---its drain term is
left in its steady form. Likewise, the dissociation wall factor $W(n)$
represents the removal of molecules at the dissociation boundary, which acts only
once the advancing population front reaches $n_{\rm diss}$. We therefore account
for transient plateau filling by introducing a dimensionless unsteady correction
factor $U(n) \in [0,1]$ into the V--V plateau term of Eq.~\eqref{eq:plateau}:
\begin{equation}
    \frac{N_n}{N} =
    U(n)\,\frac{\Gamma\,W(n)}{n+1}
    - \frac{B_{\rm VT}\,e^{\delta_{\rm VT}n}}{n+1}
    \qquad \text{for } n > n^\star,
    \label{eq:full_unsteady}
\end{equation}
where $U(n)$ is derived from the pure V--V up-pumping problem without the V--T
drain or the wall factor ($W(n)=1$). Combining $U(n)$ and $W(n)$
multiplicatively in Eq.~\eqref{eq:full_unsteady} recovers the steady AMR
distribution of Eq.~\eqref{eq:plateau} exactly as $U(n) \to 1$ and is otherwise
an approximation, assessed against StS in
Section~\ref{AccuracyOverParameterSpace}.

To derive $U(n)$, we note that the steady V--V plateau population
$N_n \propto 1/(n+1)$ varies with level, which complicates the nonlinear algebra
of the time-dependent problem. Introducing the scaled variable
\begin{equation}
    q_n \equiv (n+1)\frac{N_n}{N}
    \label{eq:qn_def}
\end{equation}
simplifies this variation: on the steady V--V plateau,
$q_n = (n+1)\cdot\Gamma/(n+1) = \Gamma$ uniformly for every $n > n^\star$, while
an empty level has $q_n = 0$. The correction factor $U(n) = q_n/\Gamma$
therefore equals unity on a fully populated level and zero on an empty one.
Treating $n$ as continuous and retaining only V--V exchange, the time evolution
of $q_n$ along the plateau is governed by the nonlinear diffusion equation of
ZNS,\cite{zhdanok1979}
\begin{equation}
    \frac{\partial q_n}{\partial t}
    = \nu(n+1)\frac{\partial^2(q_n^2)}{\partial n^2},
    \label{eq:zns}
\end{equation}
which behaves similarly to the nonlinear heat equation but in vibrational level
space, diffusing $q_n$ upward along the vibrational ladder at a rate set by the
collision frequency $\nu$.\cite{gordiets1986}

The diffusion coefficient in Eq.~\eqref{eq:zns} is
$\nu = a\,k_{\rm VV}^{1\rightarrow0,\,0\rightarrow1}\,N$, with
$a = 6hc\,\omega_{\rm e}x_{\rm e}/(k_{\rm B}T_{\rm g}\,\delta_{\rm VV}^3)$, a
dimensionless factor that follows from the V--V rate model of
Eq.~\eqref{eq:kvv_nm}. It matches the coefficient in Eq.~11 of
Ref.~\citenum{adamovich1994} apart from that equation's factor $1/{\rm e}$
(${\rm e}$ being Euler's number). That factor is not part of the V--V rate
model: it enters AMR's expression for the flux of quanta in terms of the
population at the Treanor minimum, which is not used here. Expanding
Eq.~\eqref{eq:kvv_nm} in the diffusion approximation gives a coefficient of
about 5.75, consistent with AMR's 6 without the $1/{\rm e}$. ZNS obtain a
coefficient of 4 instead of 6 because they assume a purely exponential V--V
rate, without the bracketed factor of Eq.~\eqref{eq:kvv_nm}.\cite{zhdanok1979}

Although ZNS obtained exact solutions of Eq.~\eqref{eq:zns}, those solutions
describe a steady source switched on or off at $t=0$, or a single excitation
pulse, and depend on the elapsed time since that event and on the source power
or the energy deposited by the pulse. In a repetitively pulsed discharge,
neither the elapsed time nor a single source power is defined, and because
Eq.~\eqref{eq:zns} is nonlinear, the single-pulse solution cannot be superposed
over a train of pulses. Moreover, a flow or plasma solver carries no history of
the vibrational distribution itself; it transports $e_{\rm v}$, and the rate required here is the period-averaged
material rate of change ${\rm d}e_{\rm v}/{\rm d}t$ experienced by a gas element.
In a gas at rest, this rate is computed from $e_{\rm v}$ stored at the ends of
successive interpulse intervals,
${\rm d}e_{\rm v}/{\rm d}t = f\,(e_{{\rm v},k}-e_{{\rm v},k-1})$ at the end of the
interpulse of cycle~$k$, where $f$ is the pulse repetition frequency; its
evaluation in a convective flow is described in Sec.~\ref{ModelImplementation}.

To express the transient evolution in terms of these local variables,
time is eliminated from Eq.~\eqref{eq:zns} by replacing the time derivative with
$\partial N_n/\partial t \approx N_n/\tau_{\rm e}$, where
\begin{equation}
    \tau_{\rm e} \equiv \frac{e_{\rm v}}{{\rm d}e_{\rm v}/{\rm d}t}
    \label{eq:taue}
\end{equation}
is the vibrational energy timescale, i.e., the time it would take for
$e_{\rm v}$ to reach its current value if it had always been growing at the
current period-averaged rate ${\rm d}e_{\rm v}/{\rm d}t$. This is justified
because every pulse deposits quanta into the low vibrational levels and V--V
carries them up the ladder one collision at a time, so the energy $e_{\rm v}$
grows with each cycle and serves as a record of how long the system has been
running. When $e_{\rm v}$ grows at a steady rate from zero, $\tau_{\rm e}$
equals the time elapsed since the first pulse, and as the distribution
approaches steady state, V--T losses slow the growth of $e_{\rm v}$ so that
$\tau_{\rm e} \to \infty$, recovering the steady AMR result. Replacing
$\partial/\partial t$ with $1/\tau_{\rm e}$ is an approximation, as V--V fills
the ladder from the bottom upward and different levels fill at different rates.
The level-by-level detail is nonetheless retained by the diffusion term on the
right of Eq.~\eqref{eq:zns}, which describes how V--V carries quanta up the
ladder, while only the time derivative on the left is replaced.

Since $q_n \propto N_n$, the same approximation gives
$\partial q_n/\partial t \approx q_n/\tau_{\rm e}$. Substituting into
Eq.~\eqref{eq:zns} and dividing by $\nu(n+1)$ yields
\begin{equation}
    \frac{{\rm d}^2(q_n^2)}{{\rm d}n^2} = \frac{q_n}{\nu\tau_{\rm e}\,(n+1)}.
    \label{eq:ode_pre}
\end{equation}
Recalling that $\nu = a\,k_{\rm VV}^{1\rightarrow0,\,0\rightarrow1}\,N$, the
product $\nu\tau_{\rm e}$ is naturally written as $a\Lambda$, where
\begin{equation}
    \Lambda \equiv k_{\rm VV}^{1\rightarrow0,\,0\rightarrow1}\,N\,\tau_{\rm e}
    \label{eq:Lambda}
\end{equation}
counts the number of fundamental V--V collisions per energy timescale
$\tau_{\rm e}$. Substituting $\nu\tau_{\rm e} = a\Lambda$ into
Eq.~\eqref{eq:ode_pre} gives the ODE
\begin{equation}
    \frac{{\rm d}^2(q_n^2)}{{\rm d}n^2} = \frac{q_n}{a\Lambda(n+1)},
    \label{eq:ode}
\end{equation}
in which time no longer appears.

Equation~\eqref{eq:ode} is second-order and involves an unknown upper boundary.
In a pulsed discharge, V--V exchange pushes quanta upward level by level,
requiring many collisions and therefore finite time; at any instant, a front
level $n_{\rm front}$ marks the highest level V--V has reached, above which the
distribution is empty. The front is part of the solution, introducing a third
unknown alongside the two integration constants and requiring three boundary
conditions. At $n^\star$ the solution matches the Treanor branch,
$q_{n^\star} = \Gamma$; at the front the population vanishes,
$q_{n_{\rm front}} = 0$; and the flux of quanta also vanishes at $n_{\rm front}$
since no V--V exchange can occur with the unpopulated levels above
$n_{\rm front}$, meaning ${\rm d}(q_n^2)/{\rm d}n\big|_{n_{\rm front}} = 0$. The
front $n_{\rm front}$ is treated as a continuous variable and need not be an
integer.

To reduce Eq.~\eqref{eq:ode} from second- to first-order, both sides are
multiplied by ${\rm d}(q_n^2)/{\rm d}n$. The left side becomes
${\rm d}[({\rm d}(q_n^2)/{\rm d}n)^2/2]/{\rm d}n$ by the chain rule. On the
right, we use the identity
$q_n \cdot {\rm d}(q_n^2)/{\rm d}n = 2q_n^2\,{\rm d}q_n/{\rm d}n =
(2/3)\,{\rm d}(q_n^3)/{\rm d}n$. Integrating from $n$ to $n_{\rm front}$ with
$n'$ as the dummy variable, the left evaluates by the fundamental theorem;
applying the zero-flux condition at $n_{\rm front}$, the upper limit vanishes,
leaving $-\tfrac{1}{2}({\rm d}(q_n^2)/{\rm d}n)^2$. On the right, the weight $-{\rm d}(q_{n'}^3)/{\rm d}n'$ is largest at the lower
limit $n'=n$ and decreases monotonically to zero at the front, where $q_{n'}^3$
vanishes as $(n_{\rm front}-n')^6$, as the resulting solution confirms. The
integrand is therefore concentrated near $n'=n$, and $1/(n'{+}1)$ is evaluated
there and pulled out.\cite{gordiets1986} Applying $q_{n_{\rm front}} = 0$ then gives
$\int_n^{n_{\rm front}}{\rm d}(q_{n'}^3)/{\rm d}n'\,{\rm d}n' = -q_n^3$.
Assembling and multiplying by $-2$ gives the first-order equation
\begin{equation}
    \left(\frac{{\rm d}(q_n^2)}{{\rm d}n}\right)^{\!2}
    = \frac{4q_n^3}{3\,a\Lambda(n+1)}.
    \label{eq:first_integral}
\end{equation}

Since $q_n$ decreases with $n$, we have
${\rm d}(q_n^2)/{\rm d}n = 2q_n\,{\rm d}q_n/{\rm d}n < 0$, so the negative
square root is taken. Substituting
${\rm d}(q_n^2)/{\rm d}n = 2q_n\,{\rm d}q_n/{\rm d}n$ and dividing by $2q_n$ and
then by $\sqrt{q_n}$ separates the variables:
\begin{equation}
    q_n^{-1/2}\,{\rm d}q_n = -\frac{{\rm d}n}{\sqrt{3\,a\Lambda(n+1)}}.
    \label{eq:separated}
\end{equation}
Integrating both sides from $n^\star$ to $n$, the left evaluates as
$2\sqrt{q_n} - 2\sqrt{\Gamma}$ and the right as
$-2\bigl(\!\sqrt{n+1}-\sqrt{n^\star+1}\bigr)/\sqrt{3a\Lambda}$, giving
\begin{equation}
    2\sqrt{q_n} - 2\sqrt{\Gamma}
    = -\frac{2}{\sqrt{3\,a\Lambda}}
      \Bigl(\!\sqrt{n+1} - \sqrt{n^\star+1}\Bigr).
    \label{eq:integrated}
\end{equation}
With $U(n) = q_n/\Gamma$, dividing by 2, solving for $\sqrt{q_n}$, dividing by
$\sqrt{\Gamma}$, and squaring gives
\begin{equation}
    U(n) = \left[1 -
    \frac{\sqrt{n+1}-\sqrt{n^\star+1}}{\sqrt{3\,a\Lambda\Gamma}}
    \right]^{\!2}.
    \label{eq:Un}
\end{equation}

Equation~\eqref{eq:Un} shows that the entire unsteady correction is controlled
by the single group $3a\Lambda\Gamma$. In the limit
$3a\Lambda\Gamma\to\infty$ (when
$\sqrt{3a\Lambda\Gamma} \gg \sqrt{n+1}-\sqrt{n^\star+1}$), the subtracted term
is negligible for all plateau levels and $U(n) \to 1$ uniformly, so
Eq.~\eqref{eq:full_unsteady} recovers the steady AMR distribution of
Eq.~\eqref{eq:plateau}. For finite $3a\Lambda\Gamma$, the correction $U(n)<1$
decreases with $n$, meaning high levels are always the most depleted since V--V
propagates quanta upward from the bottom of the ladder and the upper levels are
the last to fill; when $\sqrt{3a\Lambda\Gamma}$ is small, the subtracted term
reaches order unity within a few levels of $n^\star$ and the plateau collapses
from the top down.

Setting $U(n_{\rm front}) = 0$ gives the level of the distribution front,
\begin{equation}
    \sqrt{n_{\rm front}+1} = \sqrt{n^\star+1} + \sqrt{3\,a\Lambda\Gamma},
    \label{eq:nf}
\end{equation}
with $n_{\rm front}$ capped at $n_{\rm diss}$. The fraction of the plateau that has filled
follows directly from Eq.~\eqref{eq:nf}: the front advances by
$\sqrt{3a\Lambda\Gamma}$, and the plateau is completely filled when this advance
reaches $\sqrt{n_{\rm diss}+1}-\sqrt{n^\star+1}$, the distance from $n^\star$ to
the dissociation level. Defining the fill fraction as the ratio of the actual
front advance to this filled value,
\begin{equation}
    \phi \equiv \frac{\sqrt{n_{\rm front}+1} - \sqrt{n^\star+1}}{\sqrt{n_{\rm diss}+1} - \sqrt{n^\star+1}},
    \label{eq:phi_geom}
\end{equation}
where $\phi = 0$ is an empty plateau and $\phi = 1$ is fully filled.
Substituting Eq.~\eqref{eq:nf} into Eq.~\eqref{eq:phi_geom} and capping at
unity gives
\begin{equation}
    \phi = \min\!\left[1,\;\sqrt{\frac{3a\Lambda\Gamma}
    {\bigl(\!\sqrt{n_{\rm diss}+1}-\sqrt{n^\star+1}\bigr)^2}}\right].
    \label{eq:phi}
\end{equation}
Setting $\phi = 1$ shows that the plateau reaches its filled, quasi-steady
regime when
\begin{equation}
    3a\Lambda\Gamma \;\geq\; \bigl(\!\sqrt{n_{\rm diss}+1}-\sqrt{n^\star+1}\bigr)^2,
    \label{eq:filled}
\end{equation}
and remains in the transient filling phase for smaller values.

\subsection{Model Implementation}
\label{ModelImplementation}
In a flow or plasma solver, the present model reconstructs the local vibrational
distribution algebraically from four macroscopic inputs: the translational gas
temperature $T_{\rm g}$, the total molecular number density $N$, the vibrational
energy per molecule measured from the ground level,
\begin{equation}
    e_{\rm v} = \frac{\sum_{n=0}^{n_{\rm diss}-1}({\mathcal E}_n-{\mathcal E}_0)\,N_n}
                     {\sum_{n=0}^{n_{\rm diss}-1}N_n},
    \label{eq:ev_def}
\end{equation}
and its period-averaged rate of change ${\rm d}e_{\rm v}/{\rm d}t$. This rate
is the period-averaged material (Lagrangian) rate of change of $e_{\rm v}$. In a
gas at rest it equals the cycle-to-cycle difference of
Sec.~\ref{TransientCorrection}, but in a convective flow at a periodic steady
state that difference vanishes at a fixed grid point even though gas elements
passing through the discharge are still accumulating vibrational energy. There,
${\rm d}e_{\rm v}/{\rm d}t$ is instead taken as the period-averaged net source
per molecule in the solver's $e_{\rm v}$ transport equation (excitation minus
V--T and other losses), which equals ${\rm D}e_{\rm v}/{\rm D}t$ when diffusion
is negligible.

Before solving for the level populations, all quantities that do not depend on
$T_{\rm v}$ are evaluated once for the given state. Specifically, the fundamental
rate coefficients $k_{\rm VV}^{1\rightarrow0,\,0\rightarrow1}$ and
$k_{\rm VT}^{1\rightarrow0}$ are computed from Eqs.~\eqref{eq:kvv}--\eqref{eq:MW},
the rate defect parameters $\delta_{\rm VV}$ and $\delta_{\rm VT}$ from
Eqs.~\eqref{eq:delta_vv} and~\eqref{eq:delta_vt}, and the V--T drain coefficient
$B_{\rm VT}$ from Eq.~\eqref{eq:BVT}, along with the diffusion prefactor $a$. When
the vibrational energy is increasing (${\rm d}e_{\rm v}/{\rm d}t > 0$), the energy
timescale $\tau_{\rm e}$ and collision parameter $\Lambda$ are computed from
Eqs.~\eqref{eq:taue} and~\eqref{eq:Lambda}. Conversely, when
${\rm d}e_{\rm v}/{\rm d}t \leq 0$, the gas is outside the transient filling
regime; the unsteady correction is bypassed by setting $U(n) = 1$ across all
plateau levels, which recovers the steady AMR distribution of
Eq.~\eqref{eq:plateau}.

With these state parameters fixed, assembling the distribution requires
determining two remaining unknowns: the vibrational temperature $T_{\rm v}$ and
the ground-state number density $N_0$. These are obtained through a nested
bisection procedure in which an outer loop updates $T_{\rm v}$ to match the
input vibrational energy $e_{\rm v}$, and an inner loop updates $N_0$ to
conserve the total molecular number density $N$. In the outer loop, each trial
value of $T_{\rm v}$ fixes the Treanor minimum $n^\star$ via
Eq.~\eqref{eq:nstar} and determines the unnormalized Treanor population ratios
$N_n/N_0$ from Eq.~\eqref{eq:treanor} for all levels $0 \leq n \leq n^\star$.
If $n^\star \geq n_{\rm diss}-1$, the Treanor branch covers every bound level and
no plateau is built. Otherwise, because the plateau parameters $\Gamma$,
$n_{\rm front}$, and $U(n)$ depend on the absolute population at $n^\star$ while
the V--T drain term in Eq.~\eqref{eq:full_unsteady} is independent of $N_0$, the
ground-state density cannot be factored out as a simple linear normalization
constant and must instead be found iteratively in the inner bisection loop. For
each trial $N_0$, the plateau amplitude $\Gamma$ at $n^\star$ is evaluated from
Eqs.~\eqref{eq:treanor} and~\eqref{eq:Gamma}, which in turn sets the active
distribution front $n_{\rm front}$ via Eq.~\eqref{eq:nf}, capped at
$n_{\rm diss}$. Across the active plateau levels $n^\star < n \leq n_{\rm front}$,
the dissociation wall factor $W(n)$ and the unsteady correction $U(n)$ are then
computed from Eqs.~\eqref{eq:wall} and~\eqref{eq:Un}. The complete distribution
$N_n$ is assembled by combining the Treanor branch of Eq.~\eqref{eq:treanor} for
$n \leq n^\star$ with the unsteady plateau of Eq.~\eqref{eq:full_unsteady} for
$n^\star < n \leq n_{\rm front}$, while setting $N_n = 0$ above $n_{\rm front}$
and at every level from the first one at which the exponential V--T drain
exceeds the first term of Eq.~\eqref{eq:full_unsteady}. The inner bisection loop
adjusts $N_0$ until the sum of $N_n$ over all bound levels
($0 \leq n \leq n_{\rm diss}-1$) equals $N$. Once $N_0$ has converged for the
given trial $T_{\rm v}$, the corresponding vibrational energy is evaluated from
Eq.~\eqref{eq:ev_def}, and the outer bisection loop adjusts $T_{\rm v}$ until this
calculated energy matches the input $e_{\rm v}$.

\section{Validation and Baseline Models}
We assess the present model by comparing its predicted vibrational
distributions against state-to-state calculations of $\mathrm{N_2}$ under
repetitive nanosecond pulsed excitation. This section outlines the StS master
equation and test matrix (Sec.~\ref{StSReference}), followed by the baseline models and the evaluation protocol used in the comparisons (Sec.~\ref{BaselineModels}), and the use of artificial intelligence (AI) tools (Sec.~\ref{AITools}).

\subsection{State-to-State Reference and Test Matrix}
\label{StSReference}
The reference solutions are obtained by integrating the StS master equation for
the 48 bound vibrational levels of $\mathrm{N_2}$ ($n=0$--$47$) and the N-atom
density at constant gas temperature $T_{\rm g}$ and total number density
$N=P/(k_{\rm B}T_{\rm g})$. The heavy-particle kinetic mechanism includes
multi-quantum V--T relaxation and V--V exchange with up to three quanta
transferred per collision partner ($m\le 3$), together with V--V dissociation
from the top three bound levels ($n=45$--$47$) and reverse three-body
recombination obtained by detailed balance. All V--V and V--T rate coefficients
are taken from the FHO--FR theory of Adamovich.\cite{adamovich2001}
The rates are tabulated at 1~K intervals from 300 to 3000~K and evaluated at the
nearest tabulated $T_{\rm g}$; because each V--V channel is tabulated in both
directions, the master equation sums over the table with a factor $1/2$.
Electron-impact vibrational excitation and superelastic de-excitation processes
with $m\le 5$ are modeled using the Maxwellian rate fits of Laporta et
al.\cite{laporta2014_psst} at an electron temperature of $T_{\rm e}=3.4$~eV,
with de-excitation rates determined by detailed balance.

Each simulation models a train of 20 rectangular 100~ns pulses repeated at
frequency $f$, initialized from a Treanor distribution (Eq.~\eqref{eq:treanor})
at $T_{\rm v}=300$~K and the prescribed $T_{\rm g}$. To ensure a controlled
energy input per cycle, each 100~ns pulse deposits a prescribed net vibrational
energy $\varepsilon_{\rm v}$ per molecule; the constant electron density during
each pulse is adjusted by bisection so that the net electron-impact vibrational
power (excitation minus superelastic de-excitation) integrates to
$\varepsilon_{\rm v}N$ over the pulse duration. Electron processes are switched
off between pulses, and the vibrational distribution is sampled at the end of
each interpulse interval so that all analyzed states reflect purely
heavy-particle relaxation. The representative distribution comparisons in
Figs.~\ref{fig:compare_closure_temp} and~\ref{fig:compare_closure_freq} fix
$P = 1$~atm and $\varepsilon_{\rm v} = 0.02$~eV, whereas the parametric
analyses in Sections~\ref{ControllingParameter}
and~\ref{AccuracyOverParameterSpace} span all combinations of three gas
temperatures (300, 750, and 1500~K), two pressures (0.041 and 1~atm), four
repetition frequencies (1, 10, 50, and 100~kHz), and three pulse energies
(0.01, 0.02, and 0.05~eV per molecule), providing 72 operating conditions and
1440 sampled interpulse states across the 20-pulse trains.

\subsection{Baseline Models and Adapted ZNS Correction}
\label{BaselineModels}
Three algebraic models are compared against the StS reference: the
quasi-steady AMR model,\cite{adamovich1994} an adapted version of the ZNS
self-similar transient theory\cite{zhdanok1979} (denoted ``AMR~+~ZNS''), and
the present model. Whereas a coupled flow or plasma solver would obtain
$e_{\rm v}$ and ${\rm d}e_{\rm v}/{\rm d}t$ from its vibrational energy
transport equation (Sec.~\ref{ModelImplementation}), here we evaluate the
analytic models \emph{a posteriori} by supplying $e_{\rm v}$ and the
cycle-to-cycle rate ${\rm d}e_{\rm v}/{\rm d}t = f\,(e_{{\rm v},k}-e_{{\rm v},k-1})$
directly from the StS solution at the end of each cycle $k$ (with $e_{{\rm v},0}$ taken from the initial distribution). Feeding the exact
StS vibrational energy into all three models (steady AMR uses only $e_{\rm v}$) isolates errors in the predicted
distribution shape from any cumulative drift in total vibrational energy.

Constructing the AMR~+~ZNS baseline requires adapting the ZNS transient theory
so that it uses the same local macroscopic inputs as the present model. Because
Eq.~\eqref{eq:zns} is nonlinear, the ZNS single-pulse solution cannot be
superposed across a train of pulses; we instead start from the ZNS switch-on
solution for a sustained source turned on at $t=0$ (Eq.~16 of
Ref.~\citenum{zhdanok1979}), which represents continuous net vibrational energy
addition over multiple cycles. In our notation, this self-similar plateau
profile reads
\begin{equation}
    \frac{N_n}{N} = \frac{C_1}{n+1} - \frac{1}{2\int_0^t\nu\,{\rm d}t},
    \qquad\text{or}\qquad
    q_n = C_1 - \frac{n+1}{2\int_0^t\nu\,{\rm d}t},
    \label{eq:zns_orig}
\end{equation}
where $C_1$ is the steady plateau amplitude set by the external source power.
To express Eq.~\eqref{eq:zns_orig} in terms of local macroscopic variables and
embed it within the AMR framework, we match $q_n$ to the Treanor branch at
$n^\star$ by setting $C_1 = \Gamma$ and shifting the linear decay to originate
at $n^\star$ (replacing $n+1$ with $n-n^\star$), and we replace the cumulative
collision integral $\int_0^t\nu\,{\rm d}t$ with the local collision count
$\nu\tau_{\rm e} = a\Lambda$ via Eq.~\eqref{eq:taue}. Dividing the resulting
profile $q_n = \Gamma - (n-n^\star)/(2a\Lambda)$ by $\Gamma$ and flooring at
zero gives the adapted ZNS unsteady correction factor,
\begin{equation}
    U_{\rm ZNS}(n) = \max\!\left[0,\;1-\frac{n-n^\star}{2a\Lambda\Gamma}\right].
    \label{eq:Uzns}
\end{equation}
Replacing $U(n)$ with $U_{\rm ZNS}(n)$ in Eq.~\eqref{eq:full_unsteady}---while
retaining the same V--T drain and dissociation wall factor $W(n)$---ensures
that the present model and the AMR~+~ZNS baseline differ solely in the
functional form of the unsteady plateau correction, with the ZNS front
$n_{\rm front} = n^\star + 2a\Lambda\Gamma$ advancing linearly in
$a\Lambda\Gamma$ rather than as the square root of $3a\Lambda\Gamma$
(Eq.~\eqref{eq:nf}).

\subsection{Use of AI Tools}
\label{AITools}
Claude Opus~5.5 (Anthropic) was used to reduce manual derivation and
implementation effort, assisting the authors with parts of the derivations of
Secs.~\ref{SteadyState} and~\ref{TransientCorrection}, parts of the StS solver of
Sec.~\ref{StSReference}, and the scripts that produce the figures and
Table~\ref{tab:p90}. The authors verified each step of the derivations and
compared the closed-form solution with a numerical solution of
Eq.~(\ref{eq:ode}).

\section{Results and Analysis}
We evaluate the model in four steps: first by comparing its predicted
vibrational distributions directly against state-to-state calculations at
representative conditions (Sec.~\ref{TransientPlateauFilling}), next by
testing how the dimensionless group $3a\Lambda\Gamma$ controls plateau filling
(Sec.~\ref{ControllingParameter}), then by quantifying population errors across
the full parametric sweep (Sec.~\ref{AccuracyOverParameterSpace}), and finally
by examining the computational savings for multidimensional solvers
(Sec.~\ref{ComputationalCost}).

\subsection{Transient Plateau Filling}
\label{TransientPlateauFilling}
\begin{figure*}[t]
    \centering
    \includegraphics[width=0.9\textwidth]{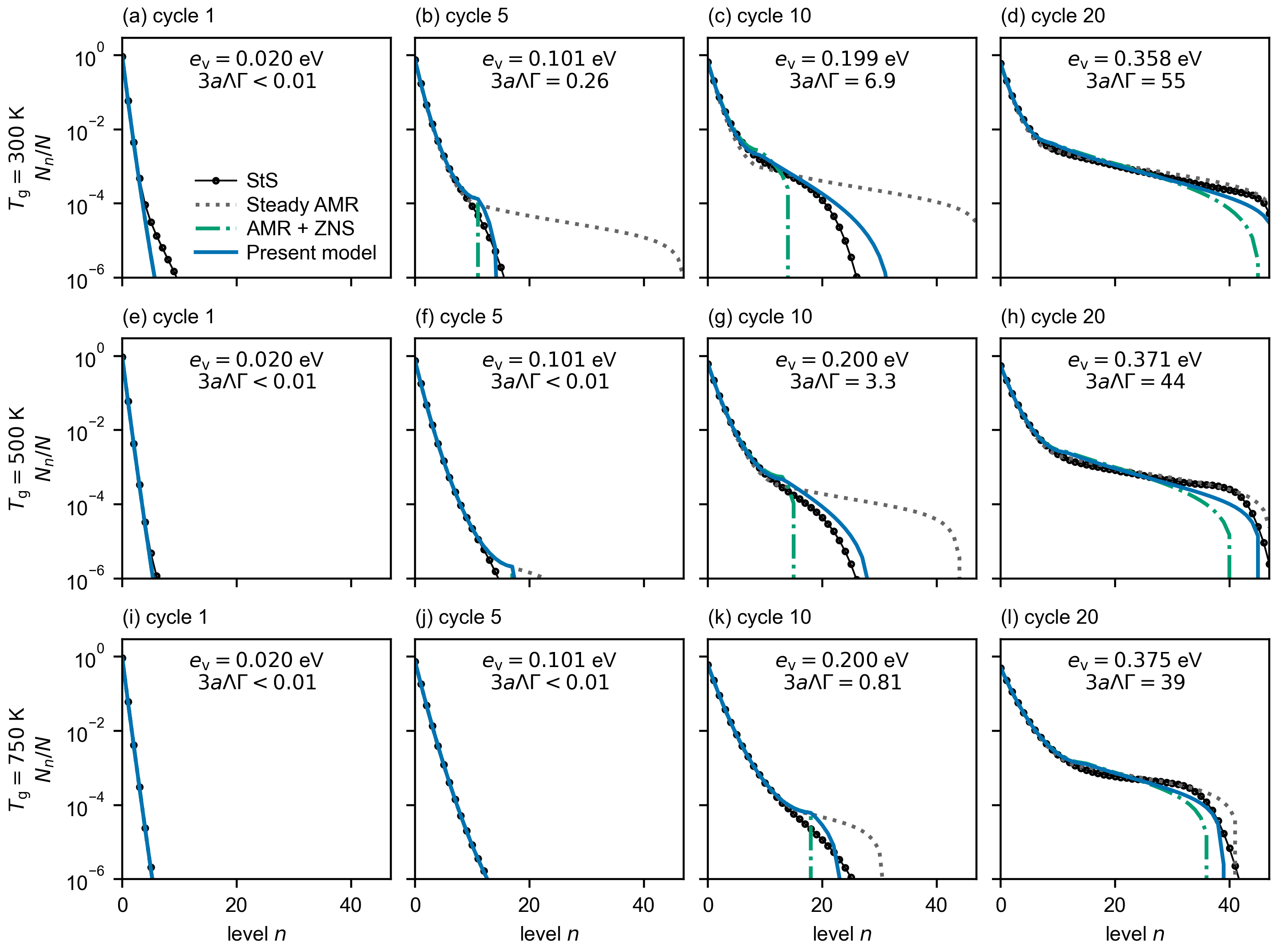}
    \figurecaption{Vibrational distributions of $\rm N_2$ at $P=1$~atm,
    sampled at the end of the interpulse of cycles 1, 5, 10, and 20,
    for 100~ns pulses repeated at $f=100$~kHz across gas temperatures
    $T_{\rm g}=300$, 500, and 750~K ($\varepsilon_{\rm v}=0.02$~eV per
    molecule per pulse). All three analytic models are evaluated from the
    same StS inputs $e_{\rm v}$ and ${\rm d}e_{\rm v}/{\rm d}t$;
    $3a\Lambda\Gamma$ is the value obtained by the present model.}
    \label{fig:compare_closure_temp}
\end{figure*}
Figure~\ref{fig:compare_closure_temp} compares the three algebraic models
against the StS reference across gas temperatures of 300, 500, and 750~K at a
pulse repetition frequency of 100~kHz. At 300~K, the plateau spans most of the
vibrational ladder, and during early cycles vibrational energy is pumped into
the low levels much faster than V--V collisions can relay quanta upward. In
this transient regime ($3a\Lambda\Gamma$ small), the three models behave quite
differently: steady AMR assumes a fully settled plateau at every cycle and
severely overpopulates the upper levels; the adapted ZNS model cuts off
prematurely just above the Treanor minimum $n^\star$; and the present model
captures the gradual, level-by-level advance of the distribution through
$U(n)$. As additional pulses accumulate---or as the gas temperature rises, which
shifts $n^\star$ upward and shortens the plateau---the distribution catches up to
its quasi-steady profile and all three models converge.

\begin{figure*}[t]
    \centering
    \includegraphics[width=0.9\textwidth]{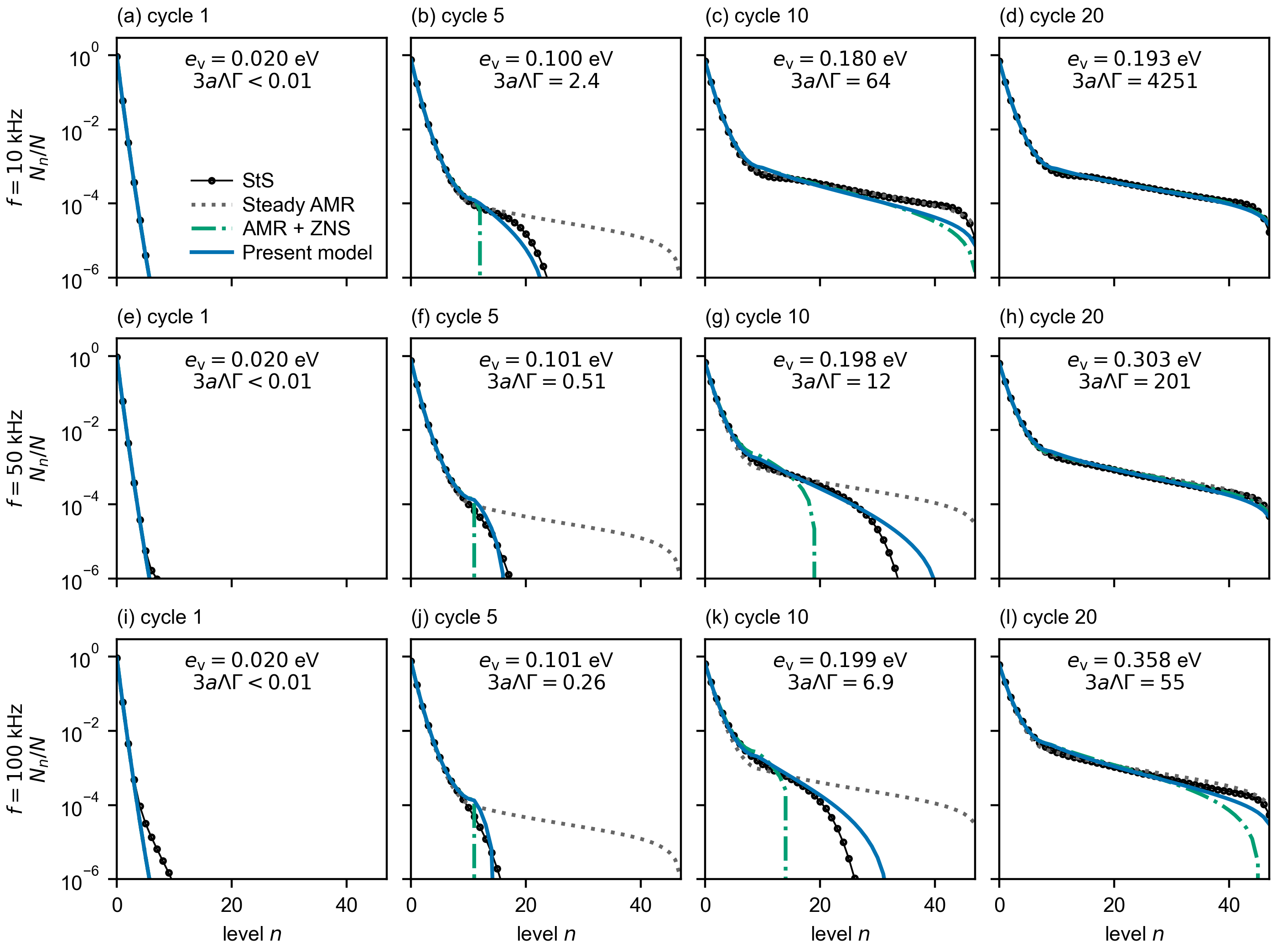}
    \figurecaption{Vibrational distributions of $\rm N_2$ at
    $T_{\rm g}=300$~K and $P=1$~atm, sampled at the end of the
    interpulse of cycles 1, 5, 10, and 20, for 100~ns pulses repeated
    at $f=10$, 50, and 100~kHz ($\varepsilon_{\rm v}=0.02$~eV per
    molecule per pulse). All three analytic models are evaluated from the
    same StS inputs $e_{\rm v}$ and ${\rm d}e_{\rm v}/{\rm d}t$;
    $3a\Lambda\Gamma$ is the value obtained by the present model.}
    \label{fig:compare_closure_freq}
\end{figure*}
Figure~\ref{fig:compare_closure_freq} fixes the gas temperature at 300~K and
steps the pulse repetition frequency across 10, 50, and 100~kHz. After the
first pulse (top row), the plateau has barely begun to form at any frequency
because the amplitude $\Gamma$ at the Treanor minimum is still vanishingly
small. Over the next several pulses, V--T losses remain negligible at 300~K, so
the accumulated vibrational energy---and hence $\Gamma$---depends mainly on how
many pulses have fired rather than on the repetition rate. What changes with
frequency is the time available between pulses for V--V collisions to do their
work: pulsing at 100~kHz leaves ten times less interpulse time for V--V
up-pumping than pulsing at 10~kHz, directly reducing $\Lambda$. As a result, at 50 and 100~kHz the plateau remains starved of quanta for roughly the first ten pulses,
accentuating both the severe overprediction of steady AMR and the premature
cutoff of the adapted ZNS model.

This early overpopulation matters most for processes that draw on the top of
the ladder, such as V--V dissociation from $n=45$--47. At 300~K, while the
plateau is still filling, StS places at most $10^{-17}$ of the molecules in
these levels, whereas steady AMR places $10^{-5}$--$10^{-4}$ there, an
overprediction of 13 to 25 orders of magnitude. The present model, like
AMR~+~ZNS, keeps its front below $n=45$ in these states and leaves the
dissociating levels empty. Once the plateau has filled, the present model and
steady AMR both give the population of these levels to within about a factor
of two of StS.

These contrasting behaviors trace directly back to the nonlinearity of V--V
transport in Eq.~\eqref{eq:zns}. Because an upward V--V step at level $n$
requires two excited molecules to collide, the upward flux of quanta scales with
the square of the local population ($q_n^2$). Near the advancing front, where
the population becomes sparse, the upward flux chokes off and causes the true
distribution to taper gradually rather than ending in a sharp cliff.

Steady AMR ignores this finite propagation speed altogether, assuming V--V
exchange has already equilibrated the entire plateau after every pulse. The
adapted ZNS model accounts for finite propagation, but its self-similar ansatz
forces the plateau correction to drop off at a constant slope
(Eq.~\eqref{eq:Uzns}) regardless of how depleted the local population has
become. Because that slope never flattens out, the ZNS distribution plunges
straight to zero just above $n^\star$, and its front advances only linearly
with $a\Lambda\Gamma$.

By solving the time-free ODE of Eq.~\eqref{eq:ode} without imposing a self-similar shape, the
present model preserves the nonlinear tapering of the V--V flux near the front.
As the population thins out at higher levels, the slope of $U(n)$ flattens
(Eq.~\eqref{eq:Un}), allowing the distribution to feather smoothly into zero
rather than cutting off abruptly. Just as importantly, the front advances as
the square root of $3a\Lambda\Gamma$ rather than linearly (Eq.~\eqref{eq:nf}).
During the first few pulses---when $3a\Lambda\Gamma$ is much smaller than
unity---its square root is much larger than the parameter itself, placing the
front many levels higher up the ladder than the linear ZNS prediction and
capturing the gradual filling seen in the state-to-state curves.

\subsection{$3a\Lambda\Gamma$ as the Controlling Parameter}
\label{ControllingParameter}
Because the distribution front advances as $\sqrt{3a\Lambda\Gamma}$
(Eq.~\eqref{eq:nf}), whether the plateau has had time to fill depends on the
combined product $3a\Lambda\Gamma$ rather than on the collision count $\Lambda$
alone. Weighting $\Lambda$ by the plateau amplitude $\Gamma$ reflects the
nonlinear nature of V--V up-pumping: a denser base at $n^\star$ supplies more
collision partners and drives the front upward much faster for the same number
of fundamental collision times.

\begin{figure*}[t]
    \centering
    \includegraphics[width=0.9\textwidth]{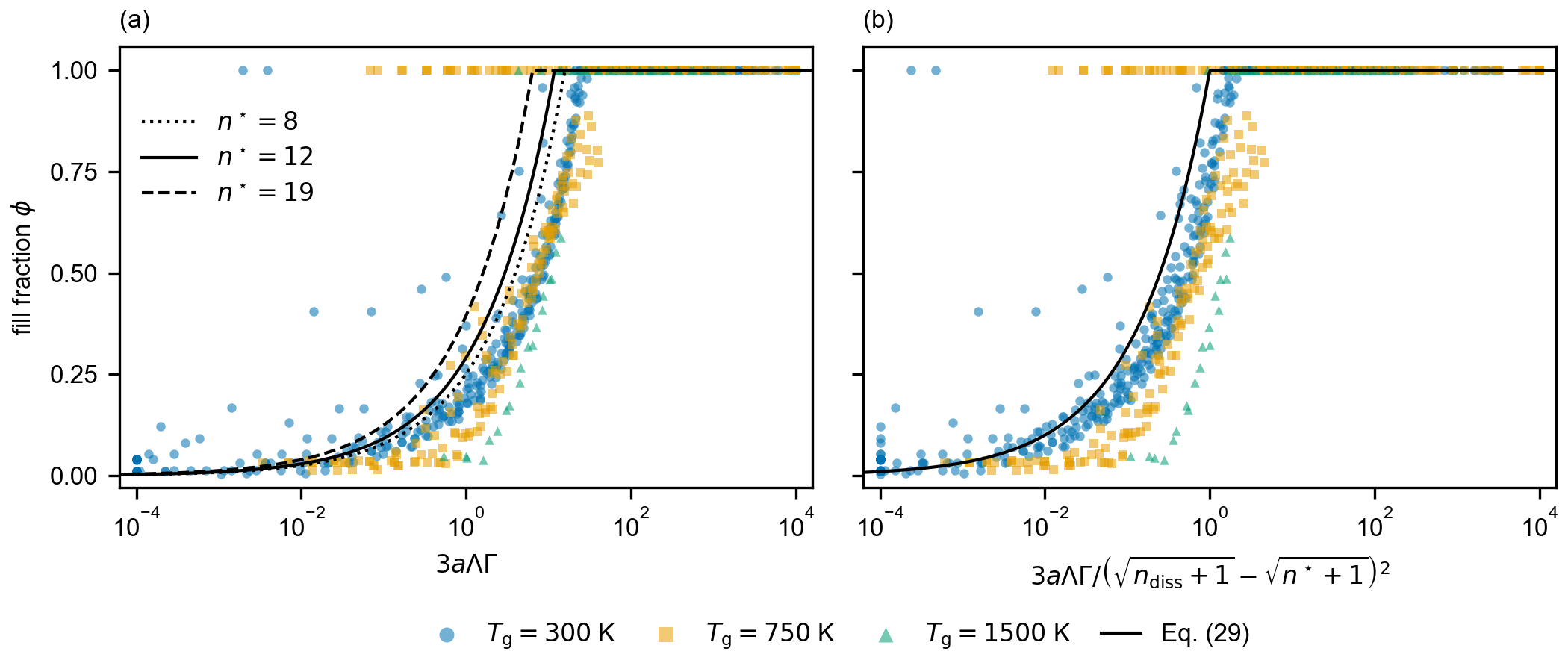}
    \figurecaption{Plateau fill fraction $\phi$ versus
    (a)~$3a\Lambda\Gamma$ and
    (b)~$3a\Lambda\Gamma/\bigl(\!\sqrt{n_{\rm diss}+1}-\sqrt{n^\star+1}\bigr)^2$
    across all plateau-bearing states in the parametric sweep (one symbol per
    interpulse interval, colored by $T_{\rm g}$). For the StS points, the
    horizontal coordinate is evaluated from the present model using the StS
    values of $e_{\rm v}$ and ${\rm d}e_{\rm v}/{\rm d}t$, while the vertical
    coordinate is obtained from Eq.~\eqref{eq:phi_geom} with $n_{\rm front}$
    defined as the first level above $n^\star$ where the StS population drops
    below $0.1$ of the steady AMR distribution at the same $e_{\rm v}$. Lines show the analytic prediction of Eq.~\eqref{eq:phi}, which
    collapses from an $n^\star$-dependent family in panel~(a) onto a single
    universal curve in panel~(b).}
    \label{fig:fill_fraction}
\end{figure*}

Figure~\ref{fig:fill_fraction} tests this scaling across all plateau-bearing
states in our parametric sweep, plotting the fraction of the plateau that has
filled ($\phi$) against the analytic prediction of Eq.~\eqref{eq:phi}. In
panel~(a), where $\phi$ is plotted against $3a\Lambda\Gamma$ directly, the
state-to-state data span several orders of magnitude on the horizontal axis yet
organize along a single rising trend. This confirms that $\Gamma$ is an
essential part of the controlling group: two conditions at different
temperatures may experience a similar number of V--V collisions $\Lambda$, but
the cooler condition has a lower Treanor minimum $n^\star$ and an
orders-of-magnitude larger plateau amplitude $\Gamma$, driving its front upward
much faster.

Because the plateau spans from $n^\star$ to $n_{\rm diss}$, the threshold value
of $3a\Lambda\Gamma$ required to reach the dissociation wall---given by
$\bigl(\!\sqrt{n_{\rm diss}+1}-\sqrt{n^\star+1}\bigr)^2$ in
Eq.~\eqref{eq:filled}---depends on $n^\star$. A higher gas temperature pushes
$n^\star$ upward, leaving a shorter plateau that takes fewer V--V collisions to
fill. Consequently, panel~(a) exhibits a family of analytic curves that
saturate at smaller $3a\Lambda\Gamma$ as $n^\star$ increases. Dividing
$3a\Lambda\Gamma$ by this filling threshold in panel~(b) removes the dependence
on plateau length, collapsing the analytic curves onto a single universal
profile where full filling occurs at unity on the horizontal axis.

While the rescaled StS data follow the shape of this universal curve in
panel~(b), two departures stand out. First, during the filling phase
($\phi < 1$), the StS points sit systematically to the right of the analytic
curve, meaning that the model's front runs slightly ahead of the StS solution---as
was visible at intermediate cycles in Figs.~\ref{fig:compare_closure_temp}(c)
and~\ref{fig:compare_closure_freq}(k). This offset arises because the model
assumes the Treanor branch below $n^\star$ is already fully equilibrated at the
instantaneous $T_{\rm v}$. In reality, V--V exchange must also build the upper
Treanor branch from the bottom up, so the population right at $n^\star$ lags
behind the equilibrium Treanor curve during rapid excitation. Reading $\Gamma$
from the equilibrium Treanor formula therefore overstates the true population
at the base of the plateau and pushes the predicted front ahead of the StS
reference. Because a higher gas temperature shifts $n^\star$ further up the
ladder onto levels that take even longer to populate, this lag---and the
resulting rightward shift in panel~(b)---grows with temperature. Eliminating this
offset would require extending the unsteady treatment to the Treanor branch
itself, which is beyond the scope of the present work. Second, the small cluster
of points lying above the curve at the far left of Fig.~\ref{fig:fill_fraction}
corresponds to initial pulses where $\Gamma$ is nearly zero; there, local V--V
collisions between the upper Treanor levels and the base of the plateau nudge
the StS front slightly past $n^\star$, a cross-branch interaction not included
in the plateau-only diffusion equation.

\subsection{Accuracy Over the Parameter Space}
\label{AccuracyOverParameterSpace}
\begin{table*}[!t]
  \center\fontsizetable
  \begin{threeparttable}
  \renewcommand{\arraystretch}{1.1}
    \tablecaption{90th percentile of the absolute error in vibrational level
    populations relative to the StS solution, in
    decades\tnote{a}~~($|\log_{10}(N_n/N_n^{\rm StS})|$), pooled over all
    populated levels and cycles of the states meeting the conditions of each
    row.}
    \label{tab:p90}
    \fontsizetable
    \begin{tabular*}{\textwidth}{@{\extracolsep{\fill}}ccccrrrr@{}}
    \toprule
    $T_{\rm g}$, K & $P$, atm & $f$, kHz & $3a\Lambda\Gamma$\tnote{b} & States\tnote{c} & Steady AMR & AMR + ZNS & Present model \\
    \midrule
    300   & all   & all  & all      &  480 & 2.21 & 1.91 & 0.46 \\
    750   & all   & all  & all      &  480 & 1.44 & 1.00 & 0.55 \\
    1500  & all   & all  & all      &  480 & 0.70 & 0.75 & 0.68 \\
    \addlinespace[1.2ex]
    all   & 0.041 & all  & all      &  720 & 2.31 & 1.79 & 0.74 \\
    all   & 1     & all  & all      &  720 & 0.82 & 0.71 & 0.33 \\
    \addlinespace[1.2ex]
    all   & all   & 1    & all      &  360 & 0.35 & 0.45 & 0.33 \\
    all   & all   & 10   & all      &  360 & 1.50 & 1.11 & 0.41 \\
    all   & all   & 50   & all      &  360 & 2.15 & 1.69 & 0.65 \\
    all   & all   & 100  & all      &  360 & 2.31 & 1.87 & 0.82 \\
    \addlinespace[1.2ex]
    all   & all   & all  & $<\bigl(\!\sqrt{n_{\rm diss}+1}-\sqrt{n^\star+1}\bigr)^2$ &  821 & 2.29 & 1.67 & 0.72 \\
    all   & all   & all  & $\ge\bigl(\!\sqrt{n_{\rm diss}+1}-\sqrt{n^\star+1}\bigr)^2$ &  619 & 0.37 & 0.82 & 0.34 \\
    \addlinespace[1.2ex]
    all   & all   & all  & all      & 1440 & 1.88 & 1.32 & 0.53 \\
    \bottomrule
    \end{tabular*}
    \begin{tablenotes}
      \item[{(a)}] A value of 1.0 corresponds to a tenfold deviation from
      the StS population, and 0.5 corresponds to a factor of
      $\sqrt{10}\approx3.2$. Levels are included when either distribution
      exceeds $10^{-6}N$, with both floored at that value.
      \item[{(b)}] Evaluated from the present model at each state; the plateau
      is filling when
      $3a\Lambda\Gamma<\bigl(\!\sqrt{n_{\rm diss}+1}-\sqrt{n^\star+1}\bigr)^2$
      and filled otherwise (Eq.~\eqref{eq:filled}).
      \item[{(c)}] ``All'' indicates pooling over every tested value of that
      parameter across the 72 operating conditions and 20-pulse trains
      ($72\times20=1440$ sampled interpulse states; see
      Sec.~\ref{StSReference}).
    \end{tablenotes}
  \end{threeparttable}
\end{table*}

Table~\ref{tab:p90} summarizes the 90th-percentile (P90) population error
across all 1440 states of the parametric sweep (considering all levels where
either distribution exceeds $10^{-6}N$). Pooled across the entire dataset
(bottom row), the steady AMR model errs by nearly two orders of magnitude
(1.88~decades) and the adapted ZNS model by more than one order of magnitude
(1.32~decades), whereas the present model brings the P90 error down to
0.53~decades---roughly a factor of three in level population.

Looking across the individual operating parameters in Table~\ref{tab:p90}, the
largest improvements consistently appear where V--V up-pumping has the furthest
to climb or the least time to get there. At 300~K, where the plateau is longest
and V--T relaxation is slow, the present model cuts the error from over two
decades for steady AMR to under half a decade (0.46). As the gas warms to
1500~K, $n^\star$ rises and rapid V--T relaxation drives the distribution to
steady state within a few pulses, so all three models converge to similar
accuracy. Lowering the pressure from 1 to 0.041~atm slows down the V--V
collision rate and keeps almost three-quarters of the sampled states in the
transient filling regime, while raising the repetition frequency toward 100~kHz
leaves too little time between pulses for the plateau to equilibrate. In both
of these strongly transient limits, steady AMR errors exceed two orders of
magnitude, whereas the present model remains below a single decade.

Splitting the dataset by the filling threshold of Eq.~\eqref{eq:filled}
confirms that these gains come entirely from the transient regime. Across the
821 states that are still actively filling, the present model reduces the error
from 2.29~decades (steady AMR) and 1.67~decades (AMR~+~ZNS) to 0.72~decades.
Once the plateau has filled (the remaining 619 states), the front has reached $n_{\rm diss}$ and the present model naturally matches steady AMR at about a third of a
decade. Notably, the adapted ZNS model actually degrades accuracy on these
filled states---as well as at 1500~K and 1~kHz---because its slow, linear front
advance continues to artificially deplete the upper plateau even when the true
distribution has already settled.

\subsection{Computational Cost}
\label{ComputationalCost}
While reducing the population error to half a decade is essential for reliable
kinetics, practical use in a multidimensional flow or plasma solver also
requires minimal per-call overhead. A coupled simulation evaluates the
vibrational distribution at every grid point on every iteration; across $10^5$
or more cells and thousands of time steps, even modest per-cell costs quickly
dominate the runtime. In a state-to-state calculation, transporting a separate
species continuity equation for every vibrational level adds 48 transport
equations, and hence 48 unknowns per grid point, for pure $\mathrm{N_2}$ and
roughly 300 for a state-resolved $\mathrm{N_2/O_2/NO/CO}$ mixture. Worse still, implicit time integration
requires assembling and factoring a chemical Jacobian across all of those
coupled levels at every step, with a factorization cost that grows as the cube
of the number of tracked vibrational states.

The present model avoids transporting individual vibrational populations
altogether. It reconstructs the distribution from four local quantities---gas
temperature, number density, vibrational energy per molecule, and its
period-averaged rate of change (obtained in a gas at rest simply by storing $e_{\rm v}$ from the end of the
previous pulse cycle, or in a convective flow from the period-averaged net
source of the $e_{\rm v}$ equation; Sec.~\ref{ModelImplementation})---all of which are already available in a
standard reacting flow or plasma solver carrying 20 to 30 macroscopic scalars.
In a multi-species plasma simulation, once the model is extended to mixtures
(Sec.~\ref{Conclusions}), replacing roughly 300 transported level
populations with a single vibrational energy equation per diatomic gas would cut the
number of unknowns per cell by an order of magnitude, shrink the Jacobian
memory footprint by two orders of magnitude, and reduce matrix factorization
cost by three orders of magnitude. At each grid point, the level populations
are then recovered locally through the simple scalar bisection loop of
Sec.~\ref{ModelImplementation} with no differential equations or matrix
operations. Because steady AMR, AMR~+~ZNS, and the present model all run this
exact same bisection loop and differ only in the closed-form factor $U(n)$
multiplying each plateau level, the present model achieves its
threefold-to-fourfold reduction in logarithmic error at essentially zero
additional computational cost.

\section{Conclusions}
\label{Conclusions}
This work has derived and validated an analytic model that extends the
quasi-steady AMR vibrational distribution to the transient plateau-filling
regime of periodically driven molecular plasmas. By replacing elapsed time in
the ZNS nonlinear V--V diffusion equation with a local vibrational energy
timescale $\tau_{\rm e} = e_{\rm v}/({\rm d}e_{\rm v}/{\rm d}t)$, the governing
PDE in level space is reduced to a time-free ODE that can be solved in closed
form without linearizing the V--V flux. The resulting unsteady correction
factor $U(n)$ multiplies the steady plateau and is controlled by a single
dimensionless group, $3a\Lambda\Gamma$, which weights the number of V--V
collisions per energy timescale by the plateau amplitude at the Treanor
minimum. Because the full nonlinearity of V--V up-pumping is retained, the
distribution front advances as the square root of $3a\Lambda\Gamma$ rather than
linearly as in adapted self-similar theory---reaching the upper ladder far
faster during early pulses, avoiding the many-order-of-magnitude overprediction of the dissociating levels by the steady model, filling the plateau once the front reaches the dissociation limit, and recovering the steady AMR plateau as $3a\Lambda\Gamma$ grows further. Across 1440 sampled states of state-to-state calculations for repetitively pulsed $\mathrm{N_2}$
discharges spanning 300--1500~K, 0.041--1~atm, and 1--100~kHz, the model
reduces the 90th-percentile population error from 1.88~decades for steady AMR
and 1.32~decades for the adapted ZNS baseline to 0.53~decades. Reconstructing
the distribution requires only four local macroscopic quantities---gas
temperature, number density, vibrational energy per molecule, and its
period-averaged rate of change---avoiding transported level populations
altogether at essentially the same evaluation cost as the steady AMR model.

Although benchmarked here for pure $\mathrm{N_2}$ under nanosecond pulse
trains, the formulation is general with respect to both gas composition and the
periodic excitation source. Because species properties enter only through
molecular, spectroscopic, and FHO constants, the model is expected to apply directly to other nonpolar diatomic gases where the AMR approximation holds. Extending
it to mixtures with inert diluents requires only AMR's mole-fraction scaling of
the V--V rate and partner-averaged V--T rates, while polar diatomics such as
$\mathrm{CO}$ require retaining the long-range V--V and radiative terms;
mixtures containing multiple V--V-active diatomics would additionally require
inter-species V--V$'$ exchange, which is not treated here. Likewise, because
the period-averaged rate ${\rm d}e_{\rm v}/{\rm d}t$ reflects only the net
accumulation of vibrational energy rather than how that energy is deposited,
the model should also apply to sinusoidal discharges and repetitive
laser-induced plasmas, although this remains to be tested.

Finally, several modeling approximations and validation bounds point to natural
directions for future work. On the theoretical side, replacing the time
derivative with $1/\tau_{\rm e}$ condenses the multi-level filling history into
a single bulk energy timescale, while the V--T drain and dissociation wall
factor are retained in their quasi-steady forms. The model also assumes that
the Treanor branch below $n^\star$ is already equilibrated at the instantaneous
vibrational temperature; as discussed in Sec.~\ref{ControllingParameter}, finite
V--V relaxation near $n^\star$ causes the upper Treanor levels to lag slightly
during rapid excitation, overstating the plateau base amplitude $\Gamma$ and
pushing the predicted front modestly ahead of the state-to-state reference. On
the validation side, the model has thus far been evaluated \emph{a posteriori}
in a stationary gas during energy build-up, where the material and
cycle-to-cycle rates of change of $e_{\rm v}$ coincide. Embedding the model
directly into a coupled multidimensional flow and plasma solver---evaluating the
material rate from the net vibrational energy source in convective flows and
extending the treatment to decaying-energy regimes---will provide a complete
assessment of its performance in predictive simulations.

\section*{Acknowledgments}
The authors are thankful to the National Science Foundation Transport Phenomena program NSF 26-518 for supporting this research under grant \#2607765. The authors are also thankful to Prof.~Igor Adamovich for fruitful discussions that helped improve the quality of the manuscript.

\section*{Data Availability}
The data that support the findings of this study are available from the corresponding author upon reasonable request.

\footnotesize
\bibliography{references}
\bibliographystyle{aiaa2}
\end{document}